\documentstyle[12pt]{article}
\newcommand{\be}{\begin{equation}}
\newcommand{\ee}{\end{equation}}
\newcommand{\bea}{\begin{eqnarray}}
\newcommand{\eea}{\end{eqnarray}}

\begin{titlepage}

\begin{flushright}
CERN-TH.6817-93
\end{flushright}
\vspace{20 mm}

\begin{center}
\huge  Super Black Holes
\vspace{1 mm}
\huge
\end{center}

\vspace{10 mm}

\begin{center}
Ulf H. Danielsson   \\
Theory Division, CERN, CH-1211 Geneva 23, Switzerland.
\end{center}

\vspace{20 mm}

\begin{center}
{\large Abstract}
\end{center}
\begin{flushleft}
A quantum version of 2D super dilaton gravity containing a black
hole is constructed for $N>8$. A previous disagreement as to whether
this is possible or not is resolved.
\end{flushright}

\vspace{10mm}
\begin{flushleft}
CERN-TH.6817/93 \\
February 1993
\end{flushleft}
\end{titlepage}
\newpage
\begin{document}
\section{Introduction}

Two-dimensional
dilaton gravity has been studied as a toy model for quantum
black holes for some time. Unfortunately there are technical
problems and ambiguities, which have so far prevented us from
obtaining a clear-cut answer to the important information puzzle,
first pointed out by Hawking \cite{hawk}. Perhaps we need to have
the full string picture, if we believe in string theory, to
resolve these issues. Another approach might be to look for
physical principles that might help us to select interesting
models. Such a principle is supersymmetry.

The supersymmetric
generalization of the original CGSH model [2--7],
has been studied in a few papers, e.g.
\cite{su1,su2,su3}.
While the classical picture is straightforward \cite{su2},
there are some
subtleties at the quantum level. The main objective of this letter is
to present a construction of a supersymmetric quantum consistent
black hole. Previously there has been a disagreement as to whether
this is possible or not \cite{su1,su3}. I will show that, although
there are restrictions due to supersymmetry, there is a way to
go around the arguments in \cite{su1}. This works, however, only
for $N>8$.

\section{Super Dilaton Gravity}

A general ansatz for the 2D super dilaton gravity action is
\be
\frac{1}{2\pi}
\int d^{2} x d^{2} \theta \left[ -iJ(\Phi ) \bar{D} D \Pi +
iK(\Phi ) \bar{D} \Phi D \Phi - \frac{i\kappa}{2} \bar{D} \Pi D \Pi +
L(\Phi , \Pi ) \right]            \label{verkan}
\ee
where $\kappa =\frac{8-N}{4}$ \cite{su2,su3}, if we assume a
trivial, flat background supermetric.
The super Liouville field and super dilaton field are given in components by
$$
\Pi = \rho + \bar{\theta} \chi +\frac{i}{2} \bar{\theta} \theta A
$$
\be
\Phi = \phi + \bar{\theta} \psi +\frac{i}{2} \bar{\theta} \theta F .
\ee
The functions $J$, $K$ and $L$ should be chosen in such a way that:
\\ \\
{\bf I}.
The weak-coupling limit of the bosonic part of the action
agrees with classical dilaton gravity.
\\ \\
{\bf II}.
The theory is an exact
superconformal field theory with vanishing central charge
(a super-string theory).
\\ \\
The latter condition is needed if
we want a quantum-mechanically-consistent theory, i.e. no dependence on
the fiducial background metric.
It anticipates the inclusion of gravity and dilaton loops, and is
certainly an important improvement on the theory. However, one should
remember, when solving the classical equations of motion, that the
resulting geometry can still only be trusted to first order in $1/N$,
which acts like $\hbar$.

To check the first condition one needs to eliminate
the auxiliary fields $A$ and $F$ using their equations
of motion. Rather than just considering the action (\ref{verkan}),
it is better
to consider a much more general case. This I will do in the next section.

\section{Eliminating Auxiliary Fields}

Let us consider the following action
\be
\frac{1}{2\pi}
\int d^{2} x d^{2} \theta \left[ i G_{\mu \nu} (\cal{X}) \bar{D} \cal{X}
^{\mu} D \cal{X} ^{\nu} +L(\cal{X}) \right]       \label{allman}
\ee
with superfields
\be
\cal{X} ^{\mu} = X^{\mu} + \bar{\theta} \psi ^{\mu} +\frac{i}{2}
\bar{\theta} \theta F^{\mu}  .
\ee
To find the bosonic part I put, for simplicity, the fermion fields to
zero. The action is then
\be
\frac{1}{2\pi}
\int d^{2} x \left[
2G_{\mu \nu} (X) \left( -\partial ^{\alpha} X^{\mu}
\partial _{\alpha} X^{\nu} +F^{\mu} F^{\nu} \right) +\partial _{\mu}
L(X) F^{\mu} \right]  \label{tjohej}   .
\ee
The equations of motion for the auxiliary fields are
\be
4G_{\mu\nu} F^{\nu} +\partial _{\mu} L =0  ,
\ee
with solution
\be
F^{\mu} = -\frac{1}{4} G^{\mu\nu} \partial _{\nu} L  .
\ee
Inserting this into (\ref{tjohej}),
the action takes the very simple form
\be
\frac{1}{2\pi}
\int d^{2} x \left[ -
2G_{\mu \nu}  \partial ^{\alpha} X^{\mu}
\partial _{\alpha} X^{\nu} -\frac{1}{8} G^{\mu\nu}
\partial _{\mu}
L \partial _{\nu} L \right]   .
\ee
In the case that interests us, i.e. (\ref{verkan}),
the metric and its inverse are
\be
{\bf G} = \left( \begin{array}{cc}
                 K & \frac{1}{2} J'  \\
                 \frac{1}{2} J' & \frac{\kappa}{2}
                \end{array}        \right) ,\;\;
{\bf G} ^{-1} = \frac{1}{\kappa K/2 -J'^{2}/4}
  \left( \begin{array}{cc}
         \frac{\kappa}{2} & -\frac{1}{2} J' \\
               - \frac{1}{2} J' & K
                \end{array}        \right)  ,
\ee
from which
it follows from this that the bosonic part of the action is
\be
\frac{1}{2\pi}
\int d^{2} x \left[ JR +2K (\nabla \phi )^{2} -\kappa (\nabla \rho )^{2}
- \frac{1}{1-\frac{2\kappa K}{J'^{2}}} \left(
\frac{K\dot{L}^{2}}{2J'^{2}} -\frac{L'\dot{L}}{2J'} +\frac{\kappa L'^{2}}
{4J'^{2}} \right) \right]
\ee
where
$L' = \frac{\partial L}{\partial \phi}$
and
$\dot{L}= \frac{\partial L}{\partial \rho}$.
This is in agreement with \cite{su3}.

How do we apply
the conditions {\bf I} and {\bf II}?
In \cite{su3} it was shown that it is
indeed possible to satisfy {\bf I}.
As an example, the supersymmetric version of the model of \cite{suss}
was considered, where
\be
J(\Phi ) = e^{-2\Phi } -\frac{\kappa}{2} \Phi \quad
{\rm and}
\quad
K(\Phi )= 2e^{-2\Phi }    .
\ee
With
\be
L= 4\lambda (
e^{ \Pi -2 \Phi } +\frac{\kappa}{4} e^{\Pi }), \label{blutt}
\ee
the potential is equal to $4\lambda ^{2} e^{2(\rho -\phi )}$
in the reduced action,
as it should be.

Condition {\bf II}, however, was applied in a way
differing from the point of view taken in this paper. In \cite{su3} the
bosonic part of the action was made to coincide with the quantum-improved
CGSH actions, where the potential is a (1,1) field and the
kinetic term (after a field redefinition) is free.
In this letter I will
apply the conformal invariance condition at the superfield
level {\it before} the auxiliary fields are integrated out.
The quantum theory will be defined from the {\it super}action using
free fields.
The superpotential is then required to
be a (1/2,1/2) field. This is {\it not} equivalent to the above
procedure.

It is
important to realize that the removal of the auxiliary fields by their
equations of motion can only be trusted at the semiclassical level,
i.e. to lowest order in $1/N$,
if we define the quantum theory using free fields.
\footnote{I would like to acknowledge discussions with
F. Bastianelli on this point.}
It is only when the auxiliary fields are zero
(as is often the case) that
one can be cavalier about this. Let us take, as
an example, the action
(\ref{allman}) with $G_{XX} =-G_{YY}=\frac{1}{4}$ and
$G_{XY} =0$ and a background charge
$Q$ for the $Y$ field. Removing the auxiliary fields  we get
\be
\frac{1}{2\pi}
\int d^{2} x \left[ \frac{1}{2}
\left( -
(\nabla X)^{2} +(\nabla Y)^{2}\right) -
\frac{1}{2} \left(
(\partial _{X} L)^{2} -(\partial _{Y} L)^{2}\right) \right]  .
\label{enkel}
\ee
If $L= e^{\alpha \cal{X}+\beta \cal{Y}}$ we get $\sim e^{2\alpha X +2\beta Y}$
f
the bosonic part. There is also a kinetic term for the fermions and
a mixed
term $e^{\alpha X + \beta Y}
(\bar{\psi} _{X} +\bar{\psi}_{Y} )(\psi _{X} + \psi _{Y})$. Clearly, the
na\"\i ve way to calculate dimensions for the bosonic part
of the reduced action fails, although it works for the mixed fermionic
term. This is simply because
$-(2\alpha )^{2}+(2\beta )(2\beta +Q) \neq
2(-\alpha ^{2}+\beta (\beta +Q))$ in general. Only in the classical
limit, i.e. as $Q \rightarrow \infty$, is the equation valid.
I will come back to this in the next section.

We conclude that the correct prescription is to start with an exact
superconformal field theory at the superfield level with vanishing
central charge. Then we might check the semiclassical
behaviour, but nothing else,
by integrating out the auxiliary fields.

If one imposes conformal invariance at the superfield level, is there still
a way to get a sensible quantum-consistent super CGSH action? According to
\cite{su1} this is not possible. Let us review their argument.

\section{Problems in Constructing SUSY Black Holes}

The action (13) can be obtained from (10)
(the same for the full superfield version)
by
a field redefinition. With
$X=X(\phi )$ and $Y= Y(\phi ,\rho )$ it is necessary to have
\be
\left\{  \begin{array}{l}
\pm \frac{1}{2} (Y'^{2} - X'^{2})=2K \\
\pm \frac{1}{2} \dot{Y} Y' = J'  \\
\pm \frac{1}{2} \dot{Y} ^{2} =-\kappa .
         \end{array}        \right.        \label{redef}
\ee
The upper (lower) sign will always correspond to $\kappa <0$
($\kappa >0$).
There is no background charge for $X$, while for $Y$ it is
$Q=\sqrt{2|\kappa |}$.
Throughout the paper I will assume a flat background metric such that
$\hat{R} =0$. Therefore the coupling to the background metric is
invisible in the action, although it affects, of course, the energy
momentum tensors etc.
In the weak coupling limit, a comparison with the CGSH model fixes the leading
terms of $J$ and $K$ to
\be
J = e^{-2\phi} +...   \quad
{\rm and}
\quad  K = 2e^{-2\phi} + ...
\ee
Hence
\be
X = -\frac{2}{Q}  e^{-2\phi} + A\phi +...,\;\;\;\;
Y = \pm \frac{2}{Q}  e^{-2\phi} + Q\rho + B\phi +...,
\ee
where $B\pm A = -Q$.

Let us now consider
a candidate for a superpotential of the form
$e^{\alpha \cal{X} +\beta \cal{Y}}$.
Clearly this is a disaster for $-\alpha \pm \beta >0$,
where the bosonic part of the potential would
blow up exponentially faster then $\frac{1}{g^{2}} \sim
e^{-2\phi}$ when the coupling $g \rightarrow 0$, i.e. $\phi \rightarrow
-\infty$.
$-\alpha \pm \beta <0$
is also unacceptable since the potential then will go to zero.
Perturbatively it is in fact identically zero.
Only the
border case $-\alpha \pm \beta =0$ retains the
possibility of a sensible leading
term. Hence we are lead to the conclusion that we must have
$\alpha = \pm \beta = \pm \frac{1}{Q}$. However, it then follows that the
potential vanishes identically in the reduced action (13). This
was the conclusion of \cite{su1}.

A slightly stronger result is obtained by proving that
\be
\left\{ \begin{array}{l}
         (\partial _{X} L)^{2} -(\partial _{Y} L)^{2} =
           \Lambda e ^{\frac{2}{Q}(X+Y)} \\
         -\partial _{X}^{2} L +\partial _{Y}^{2} L +Q\partial _{Y} L
            =L
        \end{array}   \right.           \label{pluttrix}
\ee
lacks solutions for $\Lambda \neq 0$. The general
solution of the second equation is
\be
L(u,v) =\int dx l(x) e^{xu+yv}
\label{pjutt} ,
\ee
where $u= \frac{X+Y}{2}$, $v=\frac{X-Y}{2}$ and
$-yx+\frac{Q}{2}(x-y)=1$. $l(x)$ is a density such that the integral
converges, but is otherwise arbitrary. If we insert this in the
first equation and symmetrize, we get
\be
\frac{1}{2} \int dx_{1} dx_{2} l(x_{1} ) l(x_{2} )
(x_{1} y_{2} +x_{2} y_{1} ) e^{(x_{1} +x_{2} )u+(y_{1} +y_{2} )v} =
\Lambda e^{\frac{4}{Q}u}.
\ee
The fixing of $x_{1}+x_{2}$ and $y_{1}+y_{2}$ now fixes both
$x_{1}$ and $x_{2}$ separately, unless $x_{1}+x_{2}=Q$ and
$y_{1}+y_{2}=-Q$. Hence, if we pick $x_{1}=x_{2}$, with $l(x_{1} )
\neq 0$, and find $y_{1}=y_{2} \neq 0$, we can conclude that we must
have $x_{1}y_{1} =0$, i.e. $x_{1}=0$. So, $l(x)$ can only be non-zero
for $x=0$ or $x$ such that $y=0$.
This is incompatible with the right hand side unless $\Lambda =0$.

This was a purely classical consideration. In the effective
action, however, after integrating out the fermion fields, there
will be contributions to the bosonic potential of higher order in
$\hbar$ coming from fermion loops. These will vanish for the
superpotential considered by the same reasoning as in the
classical case.

For completeness, we should check why $L$ as given by (\ref{blutt})
fails condition {\bf II}. This is especially easy in the case
\cite{suss}.
We need the field redefinitions
\be
\left\{ \begin{array}{l}
          X= -\frac{2}{Q}e^{-2\phi } -\frac{Q}{2} \phi \\
          Y= Q\rho -\frac{Q}{2} \phi +\frac{2}{Q} e^{-2\phi }
       \end{array}      \right.
\ee
(for $\kappa <0$); $L$ can be written $L=-\lambda Q e^{\rho} Y'$.
One can then show that
$\frac{\partial L}{\partial Y} = \frac{1}{Q} L$,
$\frac{\partial ^{2} L}{\partial Y^{2}} =\frac{1}{Q^{2}} L$,
$\frac{\partial L}{\partial X} = \lambda e^{\rho} X'$ and
$\frac{\partial ^{2} L}{\partial X^{2}} =
\frac{1}{Q^{2}} L +\lambda e^{\rho} \frac{X''}{X'}$. The last term
spoils the $(1/2,1/2)$ condition. In the reduced action this means that
(using free fields), the mixed bosonic--fermionic term is not (1,1).

So, it seems that we have to conclude that the theory is trivial and
that supersymmetry forbids the existence of black holes.
In the
next section a possible loop hole will be presented.

\section{A Loop Hole for SUSY Black Holes}

In order
to circumven the argument of the last section one can try to
start with
the following superpotential
\be
L= \Lambda _{1} e^{\alpha _{1} \cal{X} + \beta _{1} \cal{Y}}
+  \Lambda _{2} e^{\alpha _{2} \cal{X} + \beta _{2} \cal{Y}}  ,
\label{tjolahopp}
\ee
where
$\mp \frac{1}{2}\alpha _{i} ^{2} \pm
\frac{1}{2}\beta _{i} (\beta _{i} \pm Q) =
\frac{1}{2}$ for   $i=1,2$.
Integrating out the auxiliary fields one gets
$$
\Lambda _{1} ^{2} (\alpha _{1} ^{2} -\beta _{1}^{2} )
e^{2\alpha _{1} X +2\beta _{1} Y} +
\Lambda _{2} ^{2} (\alpha _{2} ^{2} -\beta _{2}^{2} )
e^{2\alpha _{2} X +2\beta _{2} Y}
$$
\be
+2\Lambda _{1} \Lambda _{2}
(\alpha _{1}  \alpha _{2}
-\beta _{1} \beta _{2})
e^{(\alpha _{1} +\alpha _{2})X +(\beta _{1} +\beta _{2})Y}
\label{battre}      .
\ee
The introduction of more terms than one
is by itself
not enough. Given the arguments of the previous section,
$-\alpha \pm \beta =0$ would still kill
all terms. However, the important point is that we need not demand this
for {\it all} terms. In the above case, we can arrange things so that
it is true for the mixed term, i.e. $\alpha _{1} +\alpha _{2} = \pm
(\beta _{1} +\beta _{2})$, but fails for the other two. If $-\alpha _{1}
\pm \beta _{1} <0$ one necessarily has $-\alpha _{2} \pm
\beta _{2} >0$. The
latter looks dangerous, but with
$-\alpha _{2} = \pm \beta _{2} = \pm \frac{1}{Q}$
the coefficient of the corresponding term in (\ref{battre}) vanishes,
and the bosonic part of the action is unaffected. This is only
consistent when $\pm \frac{2}{Q} >0$, that is for $\kappa <0$ or
$N>8$. Therefore we must restrict ourselves to that case.
We are left with
\be
\Lambda _{1} (\alpha _{1} ^{2} - \beta _{1}^{2}) e^{2\alpha _{1} X+
2\beta _{1} Y} +\frac{2\Lambda _{1} \Lambda _{2}}{Q} (\beta _{1}-
\alpha _{1}) e^{(\alpha _{1} -\frac{1}{Q}) X + (\beta _{1} -
\frac{1}{Q})Y}   .
\ee
We then put $\alpha _{1} -\frac{1}{Q} = \beta _{1} +\frac{1}{Q}$. The
first term is then such that it vanishes in the weak coupling limit, the
second term has the desired weak coupling behaviour and a {\it
non-vanishing} coefficient.
This can be rewritten
\be
\tilde{\Lambda} _{1} e^{\frac{2}{Q}(X-Y) +\frac{4Q}{Q^{2}-4} (X+Y)}
+\tilde{\Lambda} _{2} e^{\frac{2Q}{Q^{2}-4} (X+Y)}
\ee
or
\be
\tilde{\Lambda} _{1} e^{(\frac{4Q^{2}}{Q^{2}-4} -2)
\rho -\frac{8}{Q^{2}}e^{-2\phi} +...} +
\tilde{\Lambda} _{2} e^{\frac{2Q^{2}}{Q^{2}-4} (\rho -\phi )+...} ,
\ee
where one can check for the promised behaviour. The precise coefficient
in the last exponential is however still not quite right, it is
$\frac{2Q^{2}}{Q^{2}-4}$ rather than $2$ as in CGSH.
It becomes $2$ only in the large-$N$ limit. But as has already been
emphasized, it is only in this limit that the equations can be
trusted. The equations of motion for the auxiliary fields can
only be used in the large-$N$ limit. Hence the semiclassical
limit is the desired one.

So, I have shown that there is indeed a quantum-consistent
supersymmetric generalization of the CGSH models. After a
field redefinition, eqs.
(\ref{redef}), the superpotential is
given by (\ref{tjolahopp}), where
$$
\alpha _{1} = \frac{3Q-4/Q}{Q^{2}-4} \;\;\;\;
\beta _{1} = \frac{Q+4/Q}{Q^{2}-4}
$$
\be
\alpha _{2} = -\frac{1}{Q} \;\;\;\;
\beta _{2} = \frac{1}{Q}    .
\ee

\section{Conclusions}

We have succeeded in constructing a SUSY version of the CGSH model
which is both non-trivial and quantum-consistent. We need, however,
$N>8$ for the construction to work. Further work is needed to
understand the implications of this. One would also need to consider
the fermionic sector with more care.

\section*{Acknowledgements}

I would like to thank L. Alvarez-Gaum\'e, F. Bastianelli
and M. Kreuzer for discussions.


\begin{thebibliography}{99}

\bibitem{hawk} S. Hawking, {\it Phys. Rev.} D14 (1976) 2460.
\bibitem{CGSH} C.G. Callan, S.B. Giddings, J.A. Harvey and
A. Strominger,
               {\it Phys. Rev.} D45 (1992) R1005.
\bibitem{harstro} J. A. Harvey and
A. Strominger, "Quantum Aspects of Black Holes", Chigago preprint
EFI-92-41, hepth9209055.
\bibitem{suss} J.G. Russo, L. Susskind and  L. Thorlacius,
              {it Phys. Letters} B292  (1992) 13.
               L. Susskind and L. Thorlacius,
              {Nucl. Phys.} B382 (1992) 123.
               J.G. Russo, L. Susskind and L. Thorlacius,
              {Phys. Rev.} D46 (1992) 3444.
\bibitem{alwis}  S.P. de Alwis, "Quantum Black Holes in Two Dimensions",
Univ. Colorado preprint COLO-HEP-288, hepth9207095.
\bibitem{bilal} A. Bilal and  C.G. Callan, ``Liouville Models of Black
                Hole Evaporation", Princeton Univ. preprint
                PUPT-1320, hepth9205089.
\bibitem{stro} S. B. Giddings and A. Strominger, "Quantum Theories of
Dilaton Gravity", Santa Barbara preprint UCSBTH-92-28, hepth9207034.
\bibitem{su1} S. Nojiri and I. Oda, ``Dilaton Supergravity in Two
Dimensions and the Disappearance of Quantum Black Hole",
Japan Nat. Defense Acad. preprint
NDA-FP-8/92,
OCHA-PP-30.
\bibitem{su2} Y. Park, A. Strominger, ``Supersymmetry and Positive
Energy in Classical and Quantum Two-Dimensional Dilaton Gravity",
Santa Barbara preprint UCSBTH-92-39, hepth9210017.
\bibitem{su3} A. Bilal, ``Positive Energy Theorem and Supersymmetry
in Exactly Solvable Quantum-Corrected 2D Dilaton Gravity", Princeton
preprint PUPT-1373, hepth9301021.
\end{thebibliography}
\end{document}